%% file: pmain.tex
\newif\ifcomment
\newif\ifdraft
\newif\ifplbstyle
\newif\ifnotprltext
\newif\iflatexdiff
\latexdifftrue
\def\dvers{v0.7}
\def\srev{$Rev: 6216 $}
\def\sid{$Date: 2022-01-11 10:51:25 -0500 (Tue, 11 Jan 2022) $}
\def\dtitle{Nuclear modification factor of light neutral-meson spectra \\up to high transverse momentum in \pPb\ collisions at $\mathbf{\sqrt{s_{NN}}}$~=~8.16~TeV}
\ifplbstyle
\documentclass[preprint,times,5p,sort&compress]{elsarticle}
\usepackage{hyperref}
\newcommand{\citesupp} {\cite{suppPRL}}
\else
\documentclass[ALICE,manyauthors,11pt]{cernphprep}
\usepackage{hyperref}
\usepackage[comma,square,numbers,sort&compress]{natbib}
\newcommand{\citesupp} {\cite{suppNote}}
\fi
\usepackage{graphicx}
\usepackage{dcolumn}
\usepackage{bm} 
\usepackage{amssymb} 
\usepackage{amsfonts}
\usepackage{graphics}
\usepackage{grffile}   % to handle dots in graphics file names
\usepackage{epsfig}
\usepackage{mathrsfs}
\usepackage[loose]{units}
\usepackage[usenames,dvipsnames]{xcolor}
\usepackage[normalem]{ulem} % for strikethroughs (\sout{})
\usepackage{color}
\usepackage[utf8]{inputenc}
\usepackage[T1]{fontenc}
\usepackage{subfigure}
\usepackage{booktabs}
\iflatexdiff
\RequirePackage{color}\definecolor{RED}{rgb}{1,0,0}\definecolor{BLUE}{rgb}{0,0,1}

\fi
\input{commands.tex}
\graphicspath{{./img/}}
\ifdraft
\usepackage{lineno}
\linenumbers
\modulolinenumbers[1]
\usepackage{fancyhdr}
\fi
\begin{document}
%==========================================================%
\ifplbstyle
\title{\dtitle}
\ifdraft
\author{ALICE Collaboration \\ \today \hspace{0.3cm}\color{red}DRAFT \dvers, SVN \srev\color{black}}
\else
\author{ALICE Collaboration}
\fi
% \ifdraft
% \date{\today \hspace{0.3cm}\color{red}DRAFT \dvers, SVN \srev\color{black}}
% \else
% \date{\today}
% \fi
\else
\begin{titlepage}
\PHyear{2021}
\PHnumber{053} % required, obtained from PH
\PHdate{01 April}   % required
\title{\dtitle}
\ShortTitle{Neutral meson $R_{\rm pPb}$ up to high $\pt$}
\Collaboration{ALICE Collaboration%
         \thanks{See Appendix~\ref{app:collab} for the list of collaboration members}}
\ShortAuthor{ALICE Collaboration} % appears on left page headers, do not change
\begin{center}
\ifdraft
\today\\ \color{red}DRAFT \dvers\ \hspace{0.3cm} SVN \srev \color{black}\vspace{0.3cm}
\else
%\today
\fi
\end{center}
\fi
%==========================================================%
\begin{abstract}
Neutral pion~($\pz$) and $\eta$ meson production cross sections were measured up to unprecedentedly high transverse momenta~($\pt$) in \pPb\ collisions at \sU. %the nucleon--nucleon \coma\ \sU.
The mesons were reconstructed via their two-photon decay channel in the rapidity interval $-1.3<\ycms<0.3$ in the ranges of $0.4<\pt<200$~GeV/$c$ and $1.0<\pt<50$~GeV/$c$, respectively.
The respective nuclear modification factor~($\rpp$) is presented for $\pt$ up to of 200 and 30~GeV/$c$, where the former was achieved by extending the $\pz$ measurement in \pp\ collisions at \sUp\ using the merged cluster technique.
The values of $\rpp$ are below unity for $\pt<10$ GeV/$c$, while they are consistent with unity for $\pt>10$ GeV/$c$, leaving essentially no room for final state energy loss.
The new data provide strong constraints for nuclear parton distribution and fragmentation functions over a broad kinematic range and are compared to model predictions as well as previous results at \sA.
\end{abstract}
\ifplbstyle
\maketitle
\else
\end{titlepage}
\newpage
\setcounter{page}{2}
\fi
%==========================================================%
%==========================MAIN============================%
%==========================================================%
%%%%%%%%%%%%%%%%%%%%%%%%%%%%%%%%%%%%%%%%%%%%%%%%%%
% \ifnotprltext
\section{Introduction}
\label{sec:intro}
% \fi

Measurements of identified hadron spectra in high-energy proton--proton~(\pp)\com{ and proton--nucleus~(\pA)} collisions are well suited to constrain perturbative predictions from Quantum Chromodynamics~(QCD)~\cite{Brambilla:2014jmp}.
At large momentum transfer~($Q^2$) one relies in these perturbative QCD~(pQCD) calculations on the factorization of computable short-range parton scattering processes such as quark--quark, quark--gluon and gluon--gluon scatterings from long-range properties of QCD that need experimental input.
These non-perturbative properties are typically modeled by parton distribution functions~(PDFs), which describe the fractional-momentum~($x$) distributions of quarks and gluons within the proton\com{ in the collinear approximation}, and fragmentation functions~(FFs), which describe the fractional-momentum~($z$) distribution of quarks or gluons for hadrons of certain species.

In high-energy proton--nucleus~(\pA) collisions, nuclear effects are expected to significantly affect particle production, in particular at small $x$~\cite{Salgado:2011wc}.
Previous measurements of neutral pions and charged hadrons in \pPb\ collisions at $\snn=5.02$~TeV at the LHC~\cite{Abelev:2014dsa,Aad:2016zif,Khachatryan:2016odn,Acharya:2018hzf} indeed revealed distinct deviations from binary-scaled \pp\ collisions, confirming earlier results from deuteron--gold collisions at $\snn=0.2$~TeV at RHIC~\cite{Adler:2003ii,Adams:2003im}.
The modification at low $\pt$~($\sim$1~GeV/$c$), which is commonly attributed to nuclear shadowing, can be parameterized by nuclear parton distribution functions~(nPDFs)~\cite{Kovarik:2015cma,Eskola:2016oht}.
However, the high parton densities reached at low $\pT$~($x$ as small as $\sim5\times 10^{-4}$) make the Color Glass Condensate~(CGC) framework~\cite{Gelis:2010nm} applicable which predicts strong particle suppression due to saturation of the parton phase space in nuclei~\cite{Lappi:2013zma}.
% Recently, also parton energy loss in cold nuclear matter was shown~\cite{Arleo:2020hat} to lead to suppressed particle yields at low $\pT$, while at high $\pT$~($\sim$10~GeV/$c$) parton energy loss in hot nuclear matter may play a role~\cite{Vitev:2007ve}.
Recently, also parton energy loss in cold nuclear matter was shown~\cite{Arleo:2020hat} to lead to suppressed particle yields at low $\pT$, while the previously observed collective effects in small systems~\cite{ALICE:2014dwt,CMS:2015yux,CMS:2017kcs} also imply partonic rescatterings in hot nuclear matter to play a role~\cite{Vitev:2007ve,Huss:2020whe}.
\ifcomment *OLD*
Particle production measurements at midrapidity in proton--nucleus (\pA) collisions at LHC energies allow for the study of fundamental properties of quantum chromodynamics (QCD) at small parton fractional momentum~($x$) down to $x\sim5\times 10^{-4}$~\cite{Salgado:2011wc}.
In the Color Glass Condensate~(CGC) framework~\cite{Gelis:2010nm}, the high parton densities accessible at LHC energies are expected to provide suitable conditions to test saturation effects on parton distributions~\cite{Lappi:2013zma}.
Previous measurements of neutral pions and charged hadrons in \pPb\ collisions at $\snn=5.02$~TeV at the LHC~\cite{Abelev:2014dsa,Aad:2016zif,Khachatryan:2016odn,Acharya:2018hzf} have revealed distinct deviations from scaling by binary nucleon--nucleon collisions, confirming earlier results from deuteron--gold collisions at $\snn=0.2$~TeV at RHIC~\cite{Adler:2003ii,Adams:2003im}.
At low $\pt$~($\sim$1~GeV/$c$), the observed suppression in \pPb\ compared to \pp\ collisions is typically attributed to nuclear shadowing effects, which can be parameterized by nuclear parton distribution functions~(nPDFs)~\cite{Kovarik:2015cma,Eskola:2016oht}.
At high $\pT$~($\sim$10~GeV/$c$), measurements in \pPb\ collisions also probe energy loss in nuclear matter, which would cause particle yields to be suppressed~\cite{Vitev:2007ve}. 
 Hence, such measurements allow one to disentangle initial and final state effects with regard to the observed high-$\pt$ suppression of hadron production in \PbPb\ collisions~\cite{Qin:2015srf}.
\fi

In this letter, the nuclear modification of particle yields\com{ of inclusive hadron production in \pPb\ compared to \pp\ collisions} is quantified by 
\begin{equation}
\rpp\piet = \frac{1}{A_{\mathrm{Pb}}}
\frac{\mathrm{d}^2\sigma_{\mathrm{pPb}}}{\mathrm{d} \pt\mathrm{d} y}\bigg/\frac{\mathrm{d}^{2}\sigma_{\mathrm{pp}}}{\mathrm{d} \pt\mathrm{d} y}\,,
\label{eq:1}
\end{equation}
where $A_{\mathrm{Pb}}=208$ is the nuclear mass number of lead and $\mathrm{d}^2\sigma\piet/(\mathrm{d} \pt\mathrm{d} y)$ are the $\pz$ or $\eta$ meson cross sections
measured in \pPb\ collisions at $\snn=8.16$~ TeV and in the corresponding \pp\ reference system at $\sUp$.
The new data constrain nPDFs and FFs over a large range in $x$, and $Q^2$, including the \coma\ energy dependence based on comparisons to lower-energy data~\cite{Acharya:2018hzf}.
\ifcomment *OLD*
measured with an unprecedented $\pt$ reach in \pPb\ collisions at $\snn=8.16$~ TeV for the rapidity interval $-1.3<y<0.3$ and in the corresponding \pp\ reference system at $\sUp$.
In addition, the previous measurement of $\pz$ production in \pp\ collisions at $\sUp$~\cite{Acharya:2017tlv} was extended up to $200$~GeV/$c$.
The measurements can be used to constrain nPDFs and fragmentation functions down to low $x$ and up to high momentum transfer ($Q^2$), \com{and to determine} including the \coma\ energy dependence of nuclear effects, like shadowing or gluon saturation, \com{determined from}based on comparisons to lower-energy measurements~\cite{Acharya:2018hzf}.
\fi

%%%%%%%%%%%%%%%%%%%%%%%%%%%%%%%%%%%%%%%%%%%%%%%%%%
% \ifnotprltext
\section{Experimental setup}
\label{sec:aliceDet}
% \fi
%%%%%%%%%%%%%%%%%%%%%%%%%%%%%%%%%%%%%%%%%%%%%%%%%%
The neutral mesons were reconstructed via their two-photon decay channels~$\pz(\eta)\rightarrow\gamma\gamma$ using different reconstruction techniques provided by the various subdetector systems of ALICE~\cite{Aamodt:2008zz,Abelev:2014ffa}.
% and are corrected for contributions from weak decays.
Photons are either reconstructed using the Electromagnetic Calorimeter~(EMCal), the Photon Spectrometer (PHOS) or via the Photon Conversion Method (PCM).
The latter uses e$^+$e$^-$ pairs from conversions, which are reconstructed from tracks measured in the Inner Tracking System~(ITS)~\cite{Aamodt:2010aa} and the Time Projection Chamber~(TPC)~\cite{Aamodt:2010aa} at $|\eta|<0.9$ inside a solenoidal magnetic field of $B=0.5$~T. 
The EMCal~\cite{Abeysekara:2010ze,Allen:2010stl} is a lead-scintillator sampling electromagnetic calorimeter at a radial distance of $\unit[4.28]{m}$ from the interaction point~(IP) covering $\Delta\varphi =100^\circ$ in azimuth for $|\eta|<0.7$ in pseudorapidity during the 2012 pp data taking period. %$80^\circ < \varphi < 180^\circ$
During the \pPb\ data taking in 2016, additional modules~\cite{Allen:2010stl} were available that extended the coverage to $\Delta\varphi =107^\circ$ for $|\eta|<0.7$ and added $\Delta\varphi =60^\circ$ opposite in azimuth for $0.22< |\eta|<0.7$. %$320^\circ < \varphi < 327^\circ$ for $|\eta|<0.7$ and $260^\circ < \varphi < 320^\circ$ for $0.22< |\eta|<0.7$.
The calorimeter provides an energy resolution of $\sigma_{E}/E = 4.8\%/E \oplus 11.3\%/\sqrt{E} \oplus 1.7\%$, with $E$ in units of GeV.
In its full configuration, it consists of a total of 18240 cells of transverse size $6\times6$ cm$^2$ each.
The PHOS~\cite{Dellacasa:1999kd} is a lead tungstate electromagnetic calorimeter with\com{ nominally} 12544 channels at a distance of $\unit[4.6]{m}$ from the IP, covering $\Delta\varphi=70^\circ$ and $|\eta|<0.12$. 
Its high light yield combined with its cell size being only slightly larger than the Moli\`ere radius of $\unit[2]{cm}$ results in an energy resolution of $\sigma_{E}/E = 1.8\%/E \oplus 3.3\%/\sqrt{E} \oplus 1.1\%$.
%The ITS~\cite{Aamodt:2010aa} is a six-layer silicon detector of two layers each of silicon pixel (SPD), silicon drift (SDD) and silicon strip (SSD) detectors.
%The TPC~\cite{Aamodt:2010aa}is a large cylindrical drift detector providing a maximum of 159 reconstructed space points per track as well as particle identification via the measurement of specific energy loss d$E$/d$x$.
%The central detectors are embedded in a magnetic field of $B=0.5$~T generated by a large solenoid magnet.
%Both detectors have full azimuthal coverage around the IP.

%%%%%%%%%%%%%%%%%%%%%%%%%%%%%%%%%%%%%%%%%%%%%%%%%%
% \ifnotprltext
\section{Data samples and event selection}
\label{sec:datasamples}
% \fi
%%%%%%%%%%%%%%%%%%%%%%%%%%%%%%%%%%%%%%%%%%%%%%%%%%
The $\pPb$ data at $\sU$ were recorded in 2016. % with two beam configurations where the direction of the proton and lead beams was inverted. % after half of the data taking.
Equal magnetic rigidity for proton and Pb beams in the LHC resulted in a rapidity shift of $\Delta y_{\rm NN}=0.465$ in the direction of the proton beam between the nucleon--nucleon centre-of-mass and the laboratory reference system. 
% Due to the equal magnetic rigidity for proton and Pb beams in the LHC, the nucleon--nucleon \coma\ rapidity frame is shifted by $\Delta y=0.465$ with respect to the laboratory system ($|y_{\rm lab}|<0.8$) in the direction of the proton beam. %, leading to approximately $-1.3<\ycms<0.3$.
%The \pp\ data were recorded in 2012 at $\sUp$ and for $|y|<0.8$.
The minimum bias~(MB) event trigger required a coincidence at Level~0~(L0) of signals issued by the V0A and V0C detectors, which are two arrays of 32 scintillator tiles each covering full azimuth at $2.8<\etal<5.1$ and $-3.7<\etal<-1.7$, respectively~\cite{Abbas:2013taa}.
Additional triggers at L0 required an energy deposit above $2$~GeV for EMCal  and $4$~GeV for PHOS, in $4\times4$ adjacent cells in coincidence with the MB trigger.
Based on the L0 preselection, further hardware Level~1 triggers were issued, two for the EMCal with energy thresholds at 5.5 GeV and 8 GeV and one for the PHOS at 7 GeV. % taking into account a larger area of the EMCal, and hence increasing the trigger efficiency.
% For the data presented, total integrated luminosities of $11$~nb$^{-1}$ for \pPb\ and $657$~nb$^{-1}$ for \pp\ collisions~(recorded in 2012) were inspected.
To account for the yield enhancement of the event triggers, the trigger rejection factors ($RF$) for the EMCal triggers were estimated through an error function fit to the ratio of the cluster energy spectra in their plateau regions above the respective trigger thresholds.
A similar procedure was performed for the PHOS triggers, where $RF$ is determined on the ratio of the corrected $\pi^0$ meson spectra instead.
The trigger rejection factors from these fits are given in \Table{tab:Luminosities} for all event triggers.
For the high threshold triggers, $RF$ is obtained from the product $RF_\mathrm{EMCal-L1 low/MB}\times RF_\mathrm{EMCal-L1 high/low}$ or $RF_\mathrm{PHOS-L0/MB}\times RF_\mathrm{PHOS-L1/L0}$.
Uncertainties on $RF$ are given as combined statistical and systematic uncertainties where the latter part was determined via variations of the low $E$ fit range.
The integrated luminosities ($\mathscr{L_{\rm int}}$) of each trigger sample and for each reconstruction method were calculated based on the MB cross section of $\sigma_{\mathrm{MB}}=(2.09 (2.10)\pm0.04)$ b for the p--Pb (Pb--p) collisions~\cite{ALICE-PUBLIC-2018-002} and the respective $RF$ values as $\mathscr{L_{\rm int}}=RF\times N_{\mathrm{events}}/\sigma_{\mathrm{MB}}$ and are listed in \Table{tab:Luminosities}.% together with the respective values for the pp collision data at a \coma\ energy of $\sqrt{s}=8$~TeV recorded in 2012.
For PCM-EMC lower integrated luminosities are reported due to the lack of TPC readout in two thirds of the triggered data.
The pp collision data set at a \coma\ energy of $\sqrt{s}=8$~TeV used in this analysis was recorded in 2012 and the respective integrated luminosities and $RF$ values are listed in \Table{tab:Luminosities}.

\begin{table}
  \begin{center}
    \caption{Trigger rejection factor $RF$ and total integrated luminosities based on the individual samples for the different reconstruction methods and triggers in \pp\ collisions at $\sqrt{s}=8$ TeV and p--Pb collisions at $\sU$.
    The uncertainty associated with the determination of the MB cross section of 1.9\% for p--Pb and 2.6\% for \pp\ is not included. The value in brackets corresponds to the high luminosity minimum bias data sample where TPC tracking is not available.}
    
\ifplbstyle
  \resizebox{\linewidth}{!}{
    \begin{tabular}{lccccc}        
    &  \\ 
    \midrule
    System/Trigger &  &\multicolumn{3}{l}{$\mathscr{L_{\rm int}}$ (nb$^{-1}$)}\\
    \cmidrule(lr){1-1}\cmidrule(lr){3-6}
    p--Pb & $RF$ &(m)EMC & PCM-EMC & PCM & PHOS\\
    \midrule
    MB  & -&$0.018 (0.041)$ & $0.018$  & $0.022$ &$0.036$\\
    EMCal L1 (low) &$288\pm8$&  $0.206$ & $0.081$ &- &- \\
    EMCal L1 (high) &$991\pm29$&  $5.67$ & $1.42$ &- &-\\
    PHOS L0 &$(1.66\pm0.02)\cdot10^3$&  - & - &- &$1.68$\\
    PHOS L1 &$(1.58\pm0.03)\cdot10^4$&  - & - &- &$6.42$\\
    \midrule
    pp\\
    \midrule
    MB & -&$1.94$ & $1.94$  & $2.17$ &$1.25$\\
    EMCal/PHOS L0  &$64.6\pm1.0$&  $39.4$&$39.4$ &- &$136$ \\
    EMCal L1  &$(1.47\pm0.06)\cdot10^4$&  $606$ & $606$ &- &- \\
    \bottomrule 
    \end{tabular}
    
}
  \else
  \resizebox{0.7\linewidth}{!}{
    \begin{tabular}{lccccc}        
    &  \\ 
    \midrule
    System/Trigger &  &\multicolumn{3}{l}{$\mathscr{L_{\rm int}}$ (nb$^{-1}$)}\\
    \cmidrule(lr){1-1}\cmidrule(lr){3-6}
    p--Pb & $RF$ &(m)EMC & PCM-EMC & PCM & PHOS\\
    \midrule
    MB  & -&$0.018 (0.041)$ & $0.018$  & $0.022$ &$0.036$\\
    EMCal L1 (low) &$288\pm8$&  $0.206$ & $0.081$ &- &- \\
    EMCal L1 (high) &$991\pm29$&  $5.67$ & $1.42$ &- &-\\
    PHOS L0 &$(1.66\pm0.02)\cdot10^3$&  - & - &- &$1.68$\\
    PHOS L1 &$(1.55\pm0.04)\cdot10^4$&  - & - &- &$6.42$\\
    \midrule
    pp\\
    \midrule
    MB & -&$1.94$ & $1.94$  & $2.17$ &$1.25$\\
    EMCal/PHOS L0  &$64.6\pm1.0$&  $39.4$&$39.4$ &- &$136$ \\
    EMCal L1  &$(1.47\pm0.06)\cdot10^4$&  $606$ & $606$ &- &- \\
    \bottomrule 
    \end{tabular}
    
}
\fi
    \label{tab:Luminosities}
  \end{center}
\end{table}

%%%%%%%%%%%%%%%%%%%%%%%%%%%%%%%%%%%%%%%%%%%%%%%%%%
% \ifnotprltext
\section{Analysis}
\label{sec:ana}
% \fi
%%%%%%%%%%%%%%%%%%%%%%%%%%%%%%%%%%%%%%%%%%%%%%%%%%
Reconstructed tracks were used to determine the primary vertex of the collision, which was required to be within 10~cm from the nominal IP position along the beam direction. 
 Pileup events ($\sim$1.5\% in pp) containing multiple collisions within a \unit[300]{ns} window were rejected if more than one primary vertex was reconstructed from SPD hits or if the number of SPD clusters was not correlated with the number of track candidates.
The photon and meson reconstruction methods are analogous to those described in~\Refs{Acharya:2017tlv, Acharya:2018hzf}.
To achieve an optimal uncertainty cancellation on $\rpp$, the meson analyses were performed simultaneously for the \pPb\ and \pp\ data sets using identical methods and selections, where possible.

Photon reconstruction in the EMCal (PHOS) is based on grouping adjacent cells, with energy deposits above $E_{\rm cell}^{\rm min}=\unit[100~(20)]{MeV}$,\com{induced by electromagnetic showers,} into clusters starting with a seed cell of $E_{\rm cell}^{\rm seed}>\unit[500~(50)]{MeV}$.
The thresholds for PHOS are lower due to its better energy resolution and finer granularity. %lower amount of noise.
Photon candidates in the EMCal were required to have $|\eta_\gamma|<0.67$ and a minimum of two cells in the cluster ($N_{\rm cell}^{\rm cls}\geq2$).
In addition, clusters are required to have a primarily round shape by restricting the cluster elongation ($\lzt$~\cite{Cortese:2005qfz}) to values between $0.1$ and $0.5$.
The elongation $\lzt$ is defined as
\begin{equation}
 \sigma_\mathrm{long}^2=\frac{1}{2}\left[\sigma_{\varphi\varphi}^2+\sigma_{\eta\eta}^2+\sqrt{(\sigma_{\varphi\varphi}^2-\sigma_{\eta\eta}^2)^2+4\sigma_{\varphi\eta}^4}~\right],
 \label{eq:sigmalong}
\end{equation}
where the values of $\sigma_{ab}^2=\langle ab\rangle-\langle a\rangle\langle b\rangle$ and $\langle a \rangle = (w_\mathrm{tot})^{-1}\sum w_ia_i$ are based on the weighted cell energy compared to the cluster energy and in relative $\eta$ and $\phi$ direction to the seed cell of the cluster. 
The weighting is logarithmic with $w_i=\mathrm{max}(0,4.5+\mathrm{log}(E_i/E_\mathrm{clus}))$ where the sum of all $w_i$ equals $w_\mathrm{tot}$~\cite{Cortese:2005qfz}.
Small values of $\lzt$ denote clusters with a round shape that are primarily of photonic origin, while large values of $\lzt$ describe elongated clusters, which are primarily from hadronic sources or from overlapping showers.

In PHOS, $|\eta_\gamma|<0.12$ was required and the criteria $\lzt>0.1$ and $N_{\rm cell}^{\rm cls}\geq3$ were only applied to clusters with $E>\unit[2]{GeV}$.
Hadron and electron contamination of the photon clusters in the EMCal was removed if an associated track was found with $E/p_{\rm track}<1.75$.
The suppression of false matches with the $E/p$ veto increased the photon efficiency by up to 50\% \com{in dense environments in particular} at high $\pt$ with respect to previous measurements~\cite{Acharya:2017tlv,Acharya:2018hzf}.
Corrections for the non-linear energy response of the calorimeters were applied to the cluster energy.
For the EMCal the correction was obtained from electron test beam data and from laboratory-based measurements of the low-gain shapers in the front-end electronics. %~\cite{Allen:2009aa,Abeysekara:2010ze}.
The correction is sizeable only at low $E$~(6\% at 1 GeV) and at high $E$~(14\% at 200 GeV).
It includes a residual relative energy-and-position correction, which is applied on simulated EMCal clusters to match the $\pz$ peak position in data.
An improved description of the EMCal cluster properties in simulations was achieved by introducing a cross talk emulation within the same EMCal readout card as described in Ref.~\cite{Acharya:2019jkx}.
The resulting agreement of the $\pz$ mass peak position is better than 0.3\%  between data and simulation.
For the PHOS, the energy non-linearity was corrected by fixing the reconstructed $\pz$ mass to the nominal PDG value~\cite{Acharya:2019rum}.

Photon conversions were reconstructed by combining oppositely charged tracks, originating from a common vertex up to a radius of $180$~cm, through a secondary vertex\com{~(\Vz)} finder.
Only tracks with a TPC d$E$/d$x$ within $-3\sigma$ and $+4\sigma$ of the expected values for electrons were accepted, where $\sigma$ is the d$E$/d$x$ resolution.
Additionally, tracks with $p>0.4$ GeV/$c$ and d$E$/d$x$ up to $1\sigma$ above the expected value for pions were rejected.
For tracks with $p>3.5$ GeV/$c$, this was loosened to $0.5\sigma$.
The photon conversion selection criteria were further optimized with respect to previous measurements~\cite{Acharya:2017hyu,Acharya:2017tlv} to yield about 10\% better efficiency at similar purity.

An invariant mass ($m_{\gamma\gamma}$) technique was used for the reconstruction of neutral pions and $\eta$ mesons.
For this, $m_{\gamma\gamma}$ was calculated for all possible combinations of photon candidates per event taking either both photons reconstructed by the same method~(called PCM, EMC, and PHOS), or one photon reconstructed with PCM and one with EMC~(called PCM-EMC).
The invariant mass distributions were calculated in $\pt$ intervals of the meson candidates~(examples are shown in Ref.~\citesupp ).
For each interval, the combinatorial background, obtained from event mixing, and residual correlated background were subtracted~(see Ref.~\citesupp). 
The remaining distributions were then integrated in $\sim 3\sigma$ around the fitted mass peak position to determine the raw yields.

\begin{figure}
        \includegraphics[width=\linewidth]{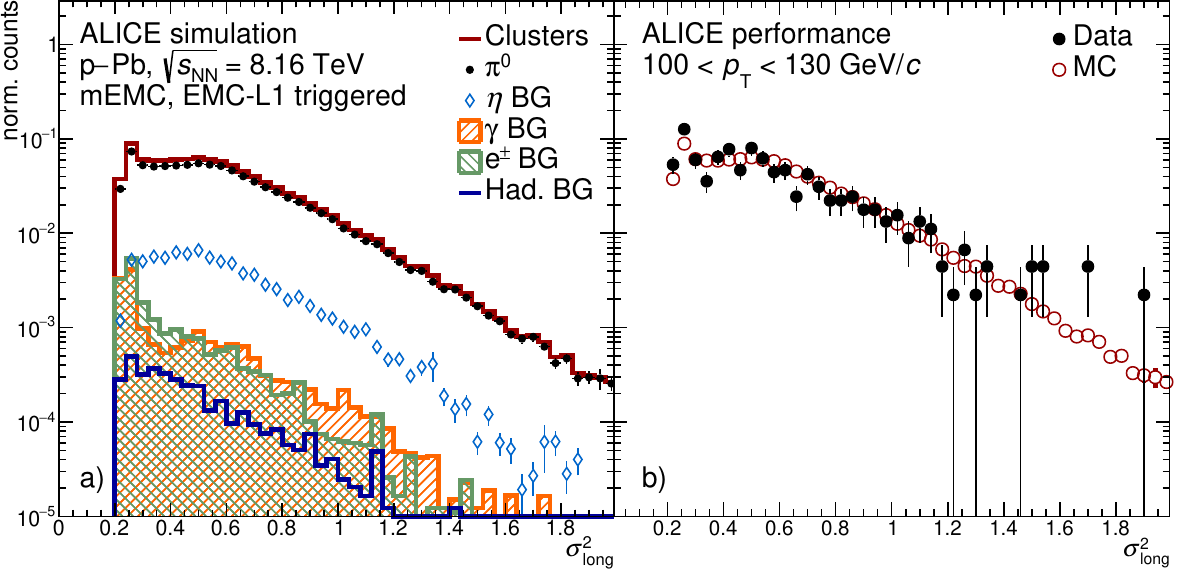}
        \caption{Shower shape distribution for the elongation $\sigma_\mathrm{long}^2$ in PYTHIA 8 Monte Carlo simulations a) showing the various contributions to the full cluster sample for a high $\pT$ example interval and b) compared to the $\sigma_\mathrm{long}^2$ distribution in data in the same $\pT$ interval.}
        \label{fig:showershape}
\end{figure}

Neutral pions with $\pt>16$~GeV/$c$ were measured with the merged-cluster~(mEMC) method~\cite{Acharya:2017hyu}, which exploits single clusters in the EMCal that result from overlapping energy deposits of both decay photons in the same cluster due to the small opening angle for large pion momentum.
The elongation~($\lzt>0.27$) of clusters with $\pt>16$~GeV/$c$ was used to discriminate between single-photon~($\lzt\approx0.25$) and merged-photon clusters.
The $\lzt$ distribution was obtained in $\pt$-intervals of the clusters and the integrated counts above $0.27$ were used as raw $\pz$ candidate yields.
An exemplary $\lzt$ distribution at high $\pt$ is shown in \Fig{fig:showershape}a in simulation, broken up into the individual contributions to the full cluster sample.
\Fig{fig:showershape}b shows a comparison between the data and simulation $\lzt$ distributions, highlighting their good agreement within uncertainties.
The resulting $\pz$ purity is between 81--87\% decreasing with $\pT$ in \pPb\ and 83--89\% in \pp\ collisions. 
It was determined via PYTHIA~8~\cite{Sjostrand:2007gs} simulations with additional data-driven corrections, which increase the relative fractions of prompt photons by 1--3\% and of $\eta$ mesons by 2\%.
POWHEG-Box~\cite{Alioli:2010xd,Alioli:2008gx} simulations were used to determine an additional purity correction for electrons from weak decays of up to 3\%.
\ifcomment % Text for 0.2:
The resulting purity is between 76(78) and 82(83)\% for \pPb~(\pp) collisions. 
It was determined via PYTHIA~8~\cite{Sjostrand:2007gs} simulations with additional data-driven corrections to the relative fraction of prompt photons~(2--4\%) and $\eta$ mesons~(2\%) as well as correction for electrons from weak decays~(up to 6\%) based on POWHEG-Box~\cite{Alioli:2010xd,Alioli:2008gx} simulations.
\fi

Correction factors for reconstruction efficiency and kinematic acceptance~(see Ref.~\citesupp) were obtained from simulations of the detector response with GEANT3~\cite{Brun:1994aa} using DPMJET~\cite{Roesler:2000he} and PYTHIA~8~\cite{Sjostrand:2007gs}  as event generators.
The correction factors for secondary $\pz$ from long-lived strange hadron decays were obtained from a particle-decay simulation based on measured spectra and are dominated by contributions from ${\rm K_{S}^{0}}$ and $\Lambda$ decays~\cite{Acharya:2018dqe,Acharya:2017tlv}.
They amount to about 1--6\% and decrease with \pt.
For the PCM method, an additional correction for out-of-bunch pileup\com{\cite{Acharya:2018dqe,Acharya:2017tlv,Abelev:2014ypa}} of 7 to 15\% decreasing with $\pt$ was applied.

The spectra were normalized by the integrated luminosities of each trigger sample and meson reconstruction method as listed in \Table{tab:Luminosities}.

The systematic uncertainties on the $\pi^0$~($\eta$) cross sections contain contributions from the yield extraction of \mbox{1--10\%}~(2--20\%) depending on the reconstruction method and $\pt$.
Further contributions from the imperfect description of the selection variables in the simulation amount to 1--4\%~(1--6\%), while the $\pt$-independent material-budget uncertainties are 4.5\% per PCM photon, 2.8\% per EMCal photon and 2\% per PHOS photon.
Uncertainties arising from the out-of-bunch pileup determination reach 3--5\% and global uncertainties on the trigger rejection factors are 2--3\%~\citesupp.
For the mEMC analysis, the largest systematic uncertainty arises from the shower overlaps in jets, which depend on the jet fragmentation and affect the $\pz$ energy resolution in the EMCal.
This uncertainty was estimated as 7--10\%, obtained from varying the particle overlaps within clusters.
The total uncertainties on the $\pz$($\eta$) cross sections are between 5(8)\% and 20(27)\% and, due to uncertainty cancellations and correlations, between 7\% and 24\% on the $\eta/\pi^0$ ratio.
For the $\rpp$, the $\pt$-independent uncertainties cancel as well as a fraction of the remaining uncertainties resulting in a total uncertainty between 4(11)\% and 25(32)\%.
A tabulated overview of the systematic uncertainty contributions for selected $\pt$-intervals is given in \Table{tab:SysErrs}.

%%%%%%%%%%%%%%%%%%%%%%%%%%%%%%%%%%%%%%%%%%%%%%%%%%
% \ifnotprltext
\section{Results}
\label{sec:results}
% \fi
%%%%%%%%%%%%%%%%%%%%%%%%%%%%%%%%%%%%%%%%%%%%%%%%%%
\begin{figure}[htb!]
        \center
        \includegraphics[width=0.9\linewidth]{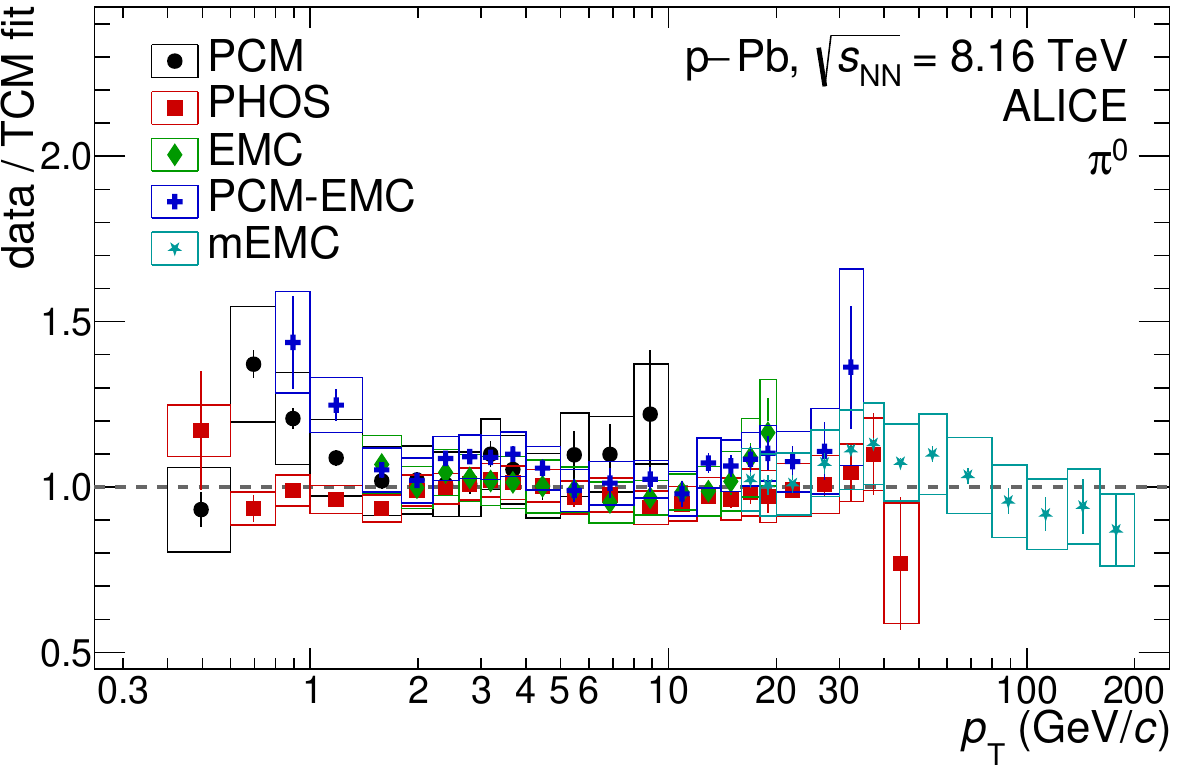}
        \caption{Ratio of the neutral pion invariant differential cross sections to the two-component model (TCM) fit of the combined spectrum for the different reconstruction techniques PCM, PCM-EMC, EMC, PHOS and mEMC in \pPb\ collisions at \sU. Statistical uncertainties are given by the vertical error bars while systematic uncertainties are shown as boxes.}
        \label{fig:indivtofit}
\end{figure}

\begin{table*}[t]
  \begin{center}
    \caption{Summary of relative systematic uncertainties in percent for selected \pT\ intervals for the $\pz$ and $\eta$ meson cross sections $\sigma_\mathrm{p-Pb}$ and nuclear modification factors $\rpp$. The statistical uncertainties are given in addition to the total systematic uncertainties for each bin. The combined statistical and systematic uncertainties, obtained by applying the BLUE method~\cite{Valassi:2013bga,Lyons:1988rp}, are also listed for all reconstruction methods available in the given \pT\ bin, considering the uncertainty correlations for the different methods. The uncertainty from the $\sigma_{\rm MB}$ determination of 1.9\%, see Ref.~\cite{ALICE-PUBLIC-2018-002}, is independent of the reported measurements and is separately indicated in the figures.}
      \resizebox{\linewidth}{!}{
      \setlength{\extrarowheight}{3pt}
    \begin{tabular}{ll llll llll lll lll lll}
     \toprule
     &Source & \multicolumn{4}{l}{$\sigma^{\pi^0}_\mathrm{p-Pb}$}& \multicolumn{4}{l}{$R_{\mathrm{pPb}}^{\pi^0}$}& \multicolumn{3}{l}{$\sigma^{\eta}_\mathrm{p-Pb}$}& \multicolumn{3}{l}{$R_{\mathrm{pPb}}^{\eta}$}& \multicolumn{3}{l}{$\eta/\pi^0_\mathrm{p-Pb}$}\\
     \cmidrule(lr){3-6}\cmidrule(lr){7-10}\cmidrule(lr){11-13}\cmidrule(lr){14-16}\cmidrule(lr){17-19}
     &$\pt$ (GeV/$c$)& 1.6 & 5.5 & 17 & 115& 1.6 & 5.5 & 17 & 115 & 2.75 & 7 & 22.5 & 2.75 & 7 & 22.5 & 2.75 & 7 & 22.5\\
     \midrule
     \textbf{PCM}
     &photon reco.          &10.7& 9.1&   -&   -&   0.7& 1.5&   -&   -&   9.5&10.5&   -&   2.0& 3.0&   -&   2.7& 4.6&   -\\
     &meson reco.           & 6.8& 6.3&   -&   -&   2.6& 7.0&   -&   -&   4.0& 5.1&   -&   4.6& 6.3&   -&   3.9& 5.2&   -\\
     &pileup                & 5.5& 3.3&   -&   -&   5.9& 6.3&   -&   -&   4.9& 4.1&   -&   6.1& 6.1&   -&   2.6& 1.9&   -\\
     &stat. uncertainty     & 2.2& 6.7&   -&   -&   3.0& 9.2&   -&   -&  12.0&25.9&   -&  18.7&37.3&   -&  12.9&25.9&   -\\
     \midrule
     \textbf{PCM-}
     &PCM photon reco.      & 4.7& 5.5& 5.0&   -&   1.3& 3.5& 3.3&   -&   7.5& 7.3& 8.1&   5.0& 5.5& 6.2&   6.0& 4.5& 7.8\\
     \textbf{EMC}&
     $\gamma$ cluster reco. & 3.5& 3.8& 4.4&   -&   1.8& 2.1& 3.8&   -&   4.8& 5.1& 6.5&   2.9& 3.3& 9.6&   3.6& 3.8& 5.0\\
     &meson reco.           & 2.4& 1.0& 1.5&   -&   2.9& 2.3& 2.2&   -&   3.3& 4.7&14.4&   5.5& 5.4& 8.3&   3.7& 6.3&14.4\\
     &trigger and efficiency& 1.0& 1.0& 3.0&   -&   0.5& 3.2& 4.9&   -&   1.0& 2.0& 3.0&   0.2& 3.2& 4.9&   1.4& 1.4& 1.4\\
     &stat. uncertainty     & 2.2& 3.7& 3.6&   -&   2.6& 5.7& 5.3&   -&  14.2&11.4&16.4&   0.0& 0.0& 0.0&  14.1&14.0&12.5\\
     \midrule
     \textbf{EMC}
     &$\gamma$ cluster reco.& 7.2& 5.9& 7.1&   -&   4.8& 2.7& 4.3&   -&   9.5& 8.5& 9.5&   7.0& 6.6& 8.2&   7.3& 6.9& 8.6\\
     &meson reco.           & 3.5& 4.1& 7.3&   -&   5.1& 4.7& 7.0&   -&  23.7& 7.3& 3.0&  29.4& 8.3& 4.8&  23.9& 7.9& 8.6\\
     &trigger and efficiency& 2.3& 2.4& 4.1&   -&   1.7& 1.7& 4.5&   -&   2.3& 3.0& 3.8&   2.6& 2.6& 5.1&   2.0& 2.5& 2.5\\
     &stat. uncertainty     & 3.1& 2.0& 4.2&   -&   4.4& 3.1& 5.3&   -&  20.8& 7.9& 6.9&  23.9&15.4&27.8&  20.9& 8.1&12.5\\
     \midrule
     \textbf{PHOS}
     &$\gamma$ cluster reco.& 3.2& 3.7& 3.8&   -&   2.0& 2.0& 2.0&   -&   4.1& 4.2& 4.2&     -&   -&   -&   0.0& 0.0& 0.0\\
     &meson reco.           & 2.1& 2.9& 4.9&   -&   3.5& 3.3& 5.2&   -&  18.0& 4.5& 7.0&     -&   -&   -&  34.6& 7.9&12.4\\
     &trigger and efficiency& 2.8& 2.8& 3.4&   -&   7.4& 7.4&14.6&   -&   1.6& 2.5& 2.5&     -&   -&   -&   1.0& 1.0& 1.0\\
     &stat. uncertainty     & 1.1& 3.1& 3.6&   -&   5.4& 6.9&11.7&   -&  29.7& 6.5&15.9&     -&   -&   -&  17.5& 5.8&12.9\\
     \midrule
     \textbf{mEMC}
     &meson PID             &   -&   -& 5.4& 5.8&     -&   -& 2.3& 3.6&     -&   -&   -&     -&   -&   -&     -&   -&   -\\
     &cluster reco.         &   -&   -& 8.3& 9.5&     -&   -& 1.0& 1.0&     -&   -&   -&     -&   -&   -&     -&   -&   -\\
     &trigger and efficiency&   -&   -& 4.1& 4.1&     -&   -& 3.8& 3.8&     -&   -&   -&     -&   -&   -&     -&   -&   -\\
     &stat. uncertainty     &   -&   -& 2.4& 5.7&     -&   -& 2.7&10.8&     -&   -&   -&     -&   -&   -&     -&   -&   -\\
     \cmidrule(lr){2-19}\morecmidrules
     \cmidrule(lr){2-19}
     &\textbf{combined syst. uncert.} 
                            & 3.6& 3.9& 4.9&11.9&   2.7& 2.9& 2.9& 5.4&   5.6& 5.3& 6.1&   5.9& 6.8&11.5&   8.6& 4.5& 8.2\\
     &\textbf{combined stat. uncert.} 
                            & 1.0& 1.8& 2.2& 5.7&   1.8& 2.7& 2.2&10.8&   8.6& 4.5& 8.2&  11.9&11.6&17.2&   5.6& 5.3& 6.1\\
     \bottomrule
    \end{tabular}
     }

    \label{tab:SysErrs}
  \end{center}
\end{table*}

\begin{figure*}[t]
        \center
        \includegraphics[width=\textwidth]{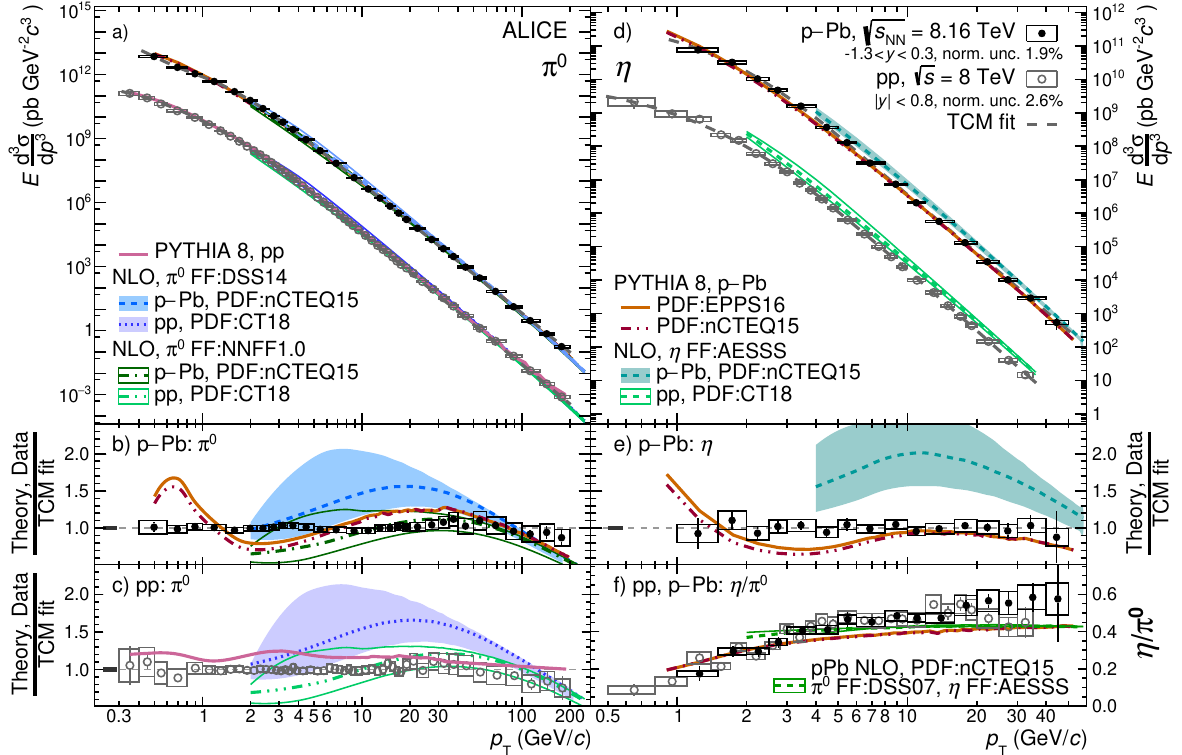}
        \caption{Neutral pion~a) and $\eta$ meson~d) \com{invariant differential} cross sections for \pp\ collisions at $\sUp$ and $\pPb$ collisions at $\sU$ together with TCM fits, NLO calculations~\cite{deFlorian:2014xna,Aidala:2010bn,Bertone:2017tyb} and PYTHIA~8~\cite{Sjostrand:2007gs,Skands:2014pea} predictions using different (n)PDFs~\cite{Kovarik:2015cma,Eskola:2016oht,Hou:2019efy}. %~(with factorization scales varied from $\pt$ to $0.5\pt$ and $2\pt$ and indicated by bands).
         Statistical uncertainties are shown as vertical bars;  the systematic uncertainties as boxes. 
         The ratios of the $\pz$ spectra in \pPb\ and \pp\ collisions to the TCM fits are shown in panel b) and c), respectively, together with the ratios of the calculations to the fits;
         panel~e) shows the same for $\eta$ mesons in \pPb\ collisions. 
         In panel f) the $\eta/\pi^0$ ratios in \pp\ and \pPb\ collisions are compared to theory predictions.
         The normalization uncertainty in the spectra ratio panels is indicated as a solid gray box around unity.}
        \label{fig:SpectrumAndRatios}
\end{figure*}

The invariant differential cross sections and \rpp\ measured by each method are consistent within their uncertainties, as shown in \Fig{fig:indivtofit}.
For the calculation of \rpp\, the spectra are shifted in the $y$-direction, while for the cross sections they are shifted along the $\pt$-axis to account for the finite bin width~\cite{Lafferty:1994cj}.
They were combined using the Best Linear Unbiased Estimate (BLUE) method~\cite{Valassi:2013bga,Lyons:1988rp} accounting for the partially correlated uncertainties. %, as described in Refs.~\cite{Acharya:2017tlv,Abelev:2014ypa,Acharya:2018hzf}.
The resulting $\pz$ and $\eta$ invariant differential cross sections for $\pPb$ collisions at $\sU$ are shown in \Fig{fig:SpectrumAndRatios} together with the $\pz$ cross section in \pp\ collisions at $\sUp$.
In both cases, the high $\pt$ reach of $\pt=200$~GeV/$c$ for the $\pz$ meson was enabled by the mEMC method, which allowed to significantly extend the previous \pp\ measurement beyond $35$~GeV/$c$~\cite{Acharya:2017tlv}.
The data is compared to a two-component model~(TCM) fit~\cite{Bylinkin:2015xya}, NLO calculations~\cite{deFlorian:2014xna,Aidala:2010bn,Bertone:2017tyb}, and PYTHIA~8~\cite{Sjostrand:2007gs,Skands:2014pea} predictions using different nPDFs~\cite{Kovarik:2015cma,Eskola:2016oht,Hou:2019efy}.
NLO calculations using the CT18~\cite{Guzzi:2011sv} PDF or nCTEQ15~\cite{Kovarik:2015cma} nPDF together with DSS14~\cite{deFlorian:2014xna} or AESSS~\cite{Aidala:2010bn} fragmentation functions generally overestimate the $\pi^0$ and $\eta$ spectra, while predicting a steeper falling spectrum at high $\pt$.
Additional NLO calculations based on the more recent NNFF1.0~\cite{Bertone:2017tyb} fragmentation functions are generally in good agreement with the data but tend to underestimate the spectra at low $\pt$.
In \Fig{fig:SpectrumAndRatios} they are shown with factorization and renormalization scales varied from $\mu=\pt$ to $\mu=0.5\pt$ and $2\pt$ and indicated by bands.
PYTHIA~8~\cite{Sjostrand:2007gs} calculations using EPPS16~\cite{Eskola:2016oht} and nCTEQ15~\cite{Kovarik:2015cma} nPDFs describe the data, however without fully capturing the shape of the $\pz$ spectra, in particular at low and intermediate $\pT$, and with a tendency to underestimate the $\eta$ spectra.
For the $\eta/\pi^0$ ratio, presented in \Fig{fig:SpectrumAndRatios}f, the differences in the shape and scale between data and calculations approximately cancel.
The ratio is rather well described by the predictions and is consistent over the full $\pt$ range between both collision systems.
For $\pt>4$~GeV/$c$, the $\eta/\pi^0$ ratio\com{ obtained from a fit of a constant,} is $C^{\eta/\pi^0}_{\rm pPb}=0.479\:\pm\:0.009(\mathrm{stat})\:\pm\:0.010(\mathrm{syst})$, consistent with the previous measurement at a lower \coma\ energy~\cite{Acharya:2018hzf} and with $C^{\eta/\pi^0}_{\mathrm{pp}}=0.473\:\pm\:0.006(\mathrm{stat})\:\pm\:0.011 (\mathrm{syst})$, the reevaluated $\eta/\pz$ ratio in \pp\ collisions at 8 TeV.

\begin{figure}[htb!]
%  \else
% \begin{figure}[h]
% \fi
        \center
        \includegraphics[width=.49\textwidth]{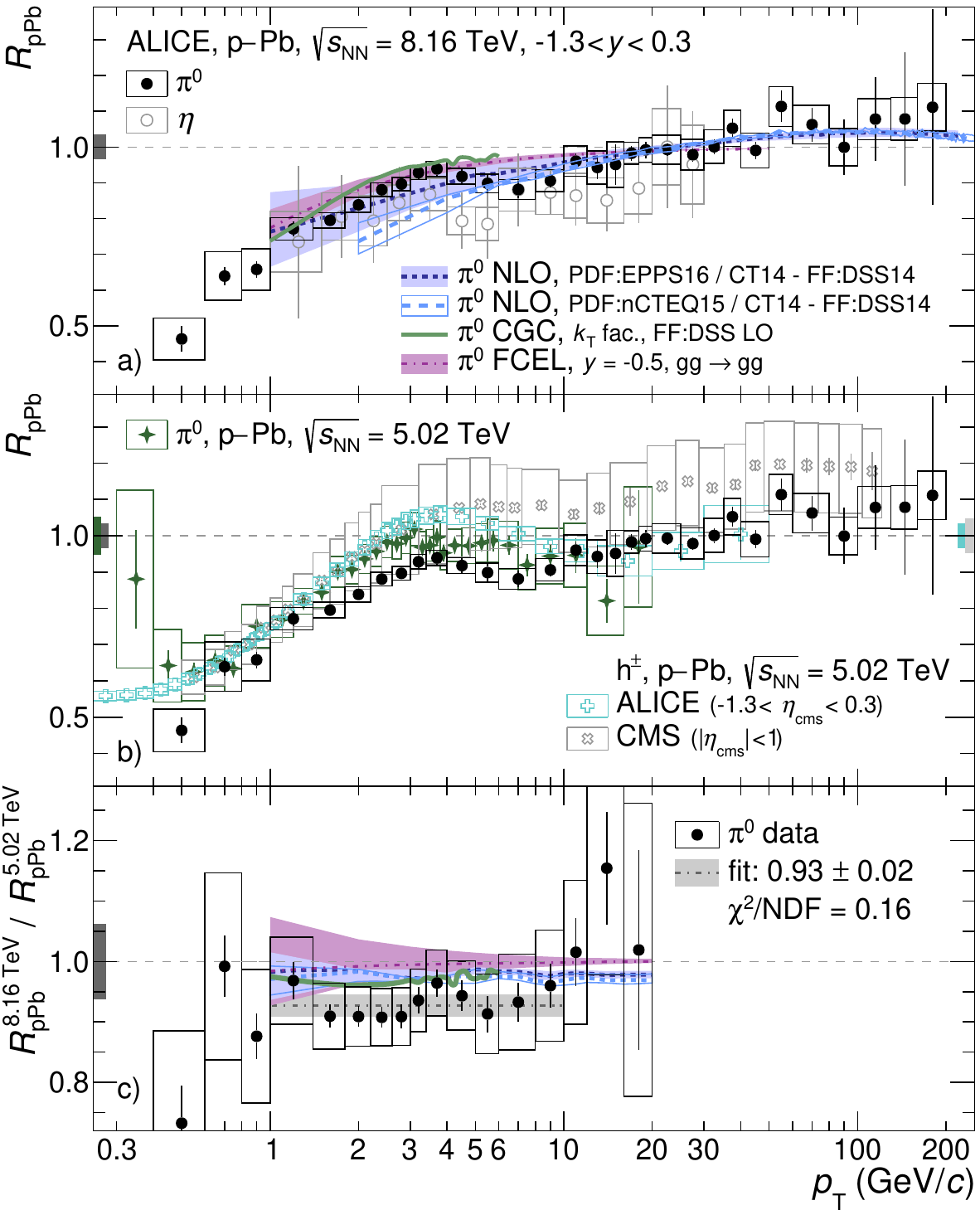}
        \caption{a)~$\rpp$ for $\pz$ and $\eta$ mesons in \pPb\  collisions at $\sU$ together with NLO~\cite{Eskola:2016oht,Kovarik:2015cma},  CGC~\cite{Lappi:2013zma} and FCEL~\cite{Arleo:2020hat} predictions.
          b)~$\rpp$ for $\pz$ at $\sU$ compared with $\pz$~\cite{Acharya:2018hzf} and charged hadron measurements~\cite{Acharya:2018qsh,Khachatryan:2016odn} at $\snn=5.02$ TeV.
          c)~Ratio of the $\pz$ $\rpp$ at $\sU$ to that at $\snn=5.02$ TeV together with corresponding CGC and FCEL model predictions.
        Statistical uncertainties are shown as vertical bars; the systematic uncertainties as boxes. 
        The overall normalization uncertainties are indicated as solid boxes around unity and amount to 3.4\% in a) and b), and to 6.2\% in c).}
        \label{fig:RpAplot}
\end{figure}

To provide the pp reference for the $\rpp$, the \pp\ spectra measured at $\sUp$ were scaled to the \pPb\ collision energy and corrected for the rapidity difference, using the ratio of $\pz$ spectra generated with PYTHIA~8 Monash 2013~\cite{Skands:2014pea} for both kinematic regions, leading to a 1--2\% increase over the whole $\pt$ range.
The resulting $\rpp$ at $\sU$ is shown in \Fig{fig:RpAplot}a for both mesons together with theory predictions and in \Fig{fig:RpAplot}b compared to data taken at $\snn=5.02$~TeV.
In the intermediate $\pt$ region, the charged particle $\rpp$ exhibits an enhancement compared to the $\pz$ data, which is historically attributed to the stronger Cronin effect for baryons~\cite{Cronin:1974zm,Rezaeian:2008ys}.
For $\pt>10$~GeV/$c$, no deviation from unity is observed within uncertainties for both mesons, consistent with predictions and the ALICE $\pz$ and h$^\pm$ measurements at $\snn=5.02$~TeV~\cite{Acharya:2018hzf,Acharya:2018qsh}, in contrast to the moderate enhancement for charged hadrons seen by the CMS experiment~\cite{Khachatryan:2016odn}.
Fitting with a constant function resulted in $1.00\pm0.01$~($0.96\pm0.04$) with a $\chi^2/\mathrm{NDF}$ of 1.04~(0.45) for the $\pz$~($\eta$) meson.
Based on the spectral slopes, the data disfavor a more than 1\% relative energy loss or an induced constant $\pt$-shift of more than 100~MeV from final-state effects in the region between 10 and 20 GeV/$c$ for both mesons, consistent with the calculations in Ref.~\cite{Huss:2020whe}. %(or a constant shift exceeding 50 MeV) 

% \ifnotprltext

For $\pt<10$~GeV/$c$, a suppression of similar magnitude is observed for both mesons within uncertainties.  %, excluding a particle species or mass dependence among the two mesons, which is also not expected from the calculations.
The suppression is described by NLO calculations using EPPS16~\cite{Eskola:2016oht} and nCETQ15~\cite{Kovarik:2015cma} nPDFs~(the latter tends to underpredict the data below 5 GeV/$c$), as well as by models using gluon recombination as the CGC-based calculations~\cite{Lappi:2013zma} or parton energy loss in cold nuclear matter in the framework of fully coherent energy loss~(FCEL)~\cite{Arleo:2020hat}.

The comparison of the $\pz$ $\rpp$ to the previous measurement at $\snn=5.02$ TeV~\cite{Acharya:2018hzf}, as shown in \Fig{fig:RpAplot}c, is consistent with unity within uncertainties, but the data hints at a stronger suppression with increasing \coma\ energy.
A stronger suppression could originate from larger shadowing in the nPDFs, which due to the smaller $x$ probed at 8.16 TeV predict a ratio of about $0.98$ in the low $\pt$ region, or from the increasing relevance of gluon saturation, as indicated by the CGC calculation~\cite{Lappi:2013zma}.
The FCEL calculation predicts a negligible difference between the two collision energies excluding coherent energy loss as the cause of a stronger suppression.
A constant fit for $\pt<10$ GeV/$c$ yields a ratio of $0.93\pm0.02_\mathrm{tot}\pm0.06_\mathrm{norm}$, where the normalization uncertainty is dominated by the interpolation of the \pz\ reference spectrum at 5.02 TeV. %where the uncertainty is dominated by the normalization uncertainty of the reference interpolation for the 5.02 TeV data.

%%%%%%%%%%%%%%%%%%%%%%%%%%%%%%%%%%%%%%%%%%%%%%%%%%
% \ifnotprltext
\section{Conclusion}
\label{sec:summary}
% \fi
%%%%%%%%%%%%%%%%%%%%%%%%%%%%%%%%%%%%%%%%%%%%%%%%%%
In summary,
\com{the $\pt$ differential invariant} cross sections for $\pz$ and $\eta$ mesons in $\pPb$ collisions at $\sU$ were measured for $0.4<\pt<200$~GeV/$c$ and $1.0<\pt<50$~GeV/$c$, respectively, providing 
constraints for nuclear parton distributions and fragmentation functions over an unprecedented kinematic range for light mesons.
By extending the reference $\pz$ measurement in \pp\ collisions at $\sUp$ to the same $\pt$ range using the mEMC method, the $\rpp$ for $\pz$ was measured up to $200$~GeV/$c$. % for the first time for an identified particle.
The $\rpp$ is consistent with unity above 10~GeV/$c$, as expected from calculations without parton energy loss, and strongly suppressed at low $\pt$, consistent with theory predictions that also include gluon shadowing or saturation effects.
\ifplbstyle
\mbox{}\\
\ifdraft
\com{
3889/3750 words including figures
3177(text: according to TeXcount sum [texcount -sum pmain.tex])
+48(E1:3*16)
+428(F1:300/(0.5*2500/(0.9*1800))+40)
+236(F2:150/(1250/(300*6))+20)
}
\fi
\fi

%==========================================================%
%==========================ENDM============================%
%==========================================================%
% \ifplbstyle
% We thank Werner Vogelsang for providing the pQCD calculations.
% \else
%%%%%%%%%%%%%%%%%%%%%%%%%%%%%%%%%%%%%%%%%%%%%%%%%%
\newenvironment{acknowledgement}{\relax}{\relax}
%%%%%%%%%%%%%%%%%%%%%%%%%%%%%%%%%%%%%%%%%%%%%%%%%%
\begin{acknowledgement}
\section*{Acknowledgments}
We thank Werner Vogelsang for providing the pQCD calculations.
\ifdraft
\else
\input{fa_2021-03-22.tex}
\fi
\end{acknowledgement}
% \fi

%%%%%%%%%%%%%%%%%%%%%%%%%%%%%%%%%%%%%%%%%%%%%%%%%%
\bibliography{pmain}
\bibliographystyle{utphys}
%%%%%%%%%%%%%%%%%%%%%%%%%%%%%%%%%%%%%%%%%%%%%%%%%%
\ifplbstyle
\else
\newpage
\appendix
%%%%%%%%%%%%%%%%%%%%%%%%%%%%%%%%%%%%%%%%%%%%%%%%%%
\section{The ALICE Collaboration}
\label{app:collab}
%%%%%%%%%%%%%%%%%%%%%%%%%%%%%%%%%%%%%%%%%%%%%%%%%%
\ifdraft
\else
\input{Alice_Authorlist_2021-03-22.tex}
\fi
\fi
%==========================================================%
\end{document}

%% file: commands.tex
\newcommand{\ycms}       {\ensuremath{y}}
\newcommand{\etal}       {\ensuremath{\eta}}
\newcommand{\pz}           {\ensuremath{\pi^{0}}}

\newcommand{\lzt}          {\ensuremath{\sigma^{2}_{\rm long}}}
\newcommand{\Vz}  	       {\ensuremath{V^{0}}}

\newcommand{\pp}           {pp}

\newcommand{\PbPb}         {\mbox{Pb--Pb}}

\newcommand{\pPb}          {\mbox{p--Pb}}
\newcommand{\pA}           {\mbox{p--A}}

\newcommand{\sUp}          {\ensuremath{\sqrt{s} = \unit[8]{TeV}}}
\newcommand{\sU}           {\ensuremath{\snn = \unit[8.16]{TeV}}}
\newcommand{\sA}           {\ensuremath{\snn = \unit[5.02]{TeV}}}
\newcommand{\pt}           {\ensuremath{p_{\mathrm{T}}}}
\newcommand{\pT}           {\pt}

\newcommand{\rpp}          {\ensuremath{R_{\mathrm{pPb}}}}
\newcommand{\snn}          {\ensuremath{\sqrt{s_{\mathrm{NN}}}}}

\newcommand{\coma}       {centre-of-mass}

\newcommand{\Fig}[1]       {Fig.~\ref{#1}}

\newcommand{\Table}[1]     {Table~\ref{#1}}
\newcommand{\Refs}[1]      {Refs.~\cite{#1}}

\newcommand{\com}[1]       {}
\def\piet{}
\iflatexdiff

\renewcommand{\xout}[1]    {\textcolor{red}{\sout{#1}}}
 
\newcommand{\old}[1]       {{\textcolor{red}{\sout{#1}}}}

\else

\renewcommand{\xout}[1]    {}
\newcommand{\old}[1]       {\relax}

\fi

%% file: fa_2021-03-22.tex
% Version: 2021-03-22

The ALICE Collaboration would like to thank all its engineers and technicians for their invaluable contributions to the construction of the experiment and the CERN accelerator teams for the outstanding performance of the LHC complex.
The ALICE Collaboration gratefully acknowledges the resources and support provided by all Grid centres and the Worldwide LHC Computing Grid (WLCG) collaboration.
The ALICE Collaboration acknowledges the following funding agencies for their support in building and running the ALICE detector:
A. I. Alikhanyan National Science Laboratory (Yerevan Physics Institute) Foundation (ANSL), State Committee of Science and World Federation of Scientists (WFS), Armenia;
Austrian Academy of Sciences, Austrian Science Fund (FWF): [M 2467-N36] and Nationalstiftung f\"{u}r Forschung, Technologie und Entwicklung, Austria;
Ministry of Communications and High Technologies, National Nuclear Research Center, Azerbaijan;
Conselho Nacional de Desenvolvimento Cient\'{\i}fico e Tecnol\'{o}gico (CNPq), Financiadora de Estudos e Projetos (Finep), Funda\c{c}\~{a}o de Amparo \`{a} Pesquisa do Estado de S\~{a}o Paulo (FAPESP) and Universidade Federal do Rio Grande do Sul (UFRGS), Brazil;
Ministry of Education of China (MOEC) , Ministry of Science \& Technology of China (MSTC) and National Natural Science Foundation of China (NSFC), China;
Ministry of Science and Education and Croatian Science Foundation, Croatia;
Centro de Aplicaciones Tecnol\'{o}gicas y Desarrollo Nuclear (CEADEN), Cubaenerg\'{\i}a, Cuba;
Ministry of Education, Youth and Sports of the Czech Republic, Czech Republic;
The Danish Council for Independent Research | Natural Sciences, the VILLUM FONDEN and Danish National Research Foundation (DNRF), Denmark;
Helsinki Institute of Physics (HIP), Finland;
Commissariat \`{a} l'Energie Atomique (CEA) and Institut National de Physique Nucl\'{e}aire et de Physique des Particules (IN2P3) and Centre National de la Recherche Scientifique (CNRS), France;
Bundesministerium f\"{u}r Bildung und Forschung (BMBF) and GSI Helmholtzzentrum f\"{u}r Schwerionenforschung GmbH, Germany;
General Secretariat for Research and Technology, Ministry of Education, Research and Religions, Greece;
National Research, Development and Innovation Office, Hungary;
Department of Atomic Energy Government of India (DAE), Department of Science and Technology, Government of India (DST), University Grants Commission, Government of India (UGC) and Council of Scientific and Industrial Research (CSIR), India;
Indonesian Institute of Science, Indonesia;
Istituto Nazionale di Fisica Nucleare (INFN), Italy;
Institute for Innovative Science and Technology , Nagasaki Institute of Applied Science (IIST), Japanese Ministry of Education, Culture, Sports, Science and Technology (MEXT) and Japan Society for the Promotion of Science (JSPS) KAKENHI, Japan;
Consejo Nacional de Ciencia (CONACYT) y Tecnolog\'{i}a, through Fondo de Cooperaci\'{o}n Internacional en Ciencia y Tecnolog\'{i}a (FONCICYT) and Direcci\'{o}n General de Asuntos del Personal Academico (DGAPA), Mexico;
Nederlandse Organisatie voor Wetenschappelijk Onderzoek (NWO), Netherlands;
The Research Council of Norway, Norway;
Commission on Science and Technology for Sustainable Development in the South (COMSATS), Pakistan;
Pontificia Universidad Cat\'{o}lica del Per\'{u}, Peru;
Ministry of Education and Science, National Science Centre and WUT ID-UB, Poland;
Korea Institute of Science and Technology Information and National Research Foundation of Korea (NRF), Republic of Korea;
Ministry of Education and Scientific Research, Institute of Atomic Physics and Ministry of Research and Innovation and Institute of Atomic Physics, Romania;
Joint Institute for Nuclear Research (JINR), Ministry of Education and Science of the Russian Federation, National Research Centre Kurchatov Institute, Russian Science Foundation and Russian Foundation for Basic Research, Russia;
Ministry of Education, Science, Research and Sport of the Slovak Republic, Slovakia;
National Research Foundation of South Africa, South Africa;
Swedish Research Council (VR) and Knut \& Alice Wallenberg Foundation (KAW), Sweden;
European Organization for Nuclear Research, Switzerland;
Suranaree University of Technology (SUT), National Science and Technology Development Agency (NSDTA) and Office of the Higher Education Commission under NRU project of Thailand, Thailand;
Turkish Atomic Energy Agency (TAEK), Turkey;
National Academy of  Sciences of Ukraine, Ukraine;
Science and Technology Facilities Council (STFC), United Kingdom;
National Science Foundation of the United States of America (NSF) and United States Department of Energy, Office of Nuclear Physics (DOE NP), United States of America.

%% file: Alice_Authorlist_2021-03-22.tex
\small
\begin{flushleft} 

S.~Acharya$^{\rm 142}$, 
D.~Adamov\'{a}$^{\rm 97}$, 
A.~Adler$^{\rm 75}$, 
J.~Adolfsson$^{\rm 82}$, 
G.~Aglieri Rinella$^{\rm 35}$, 
M.~Agnello$^{\rm 31}$, 
N.~Agrawal$^{\rm 55}$, 
Z.~Ahammed$^{\rm 142}$, 
S.~Ahmad$^{\rm 16}$, 
S.U.~Ahn$^{\rm 77}$
% ...

\bigskip

\bigskip 

\textbf{\Large Affiliation Notes}

\bigskip 

% ...
$^{\rm V}$ Also at: Institute of Theoretical Physics, University of Wroclaw, Poland\\

\bigskip

\bigskip 

\textbf{\Large Collaboration Institutes}

\bigskip 

% ...
$^{16}$ Department of Physics, Aligarh Muslim University, Aligarh, India\\
% ...
$^{31}$ Dipartimento DISAT del Politecnico and Sezione INFN, Turin, Italy\\
% ...
$^{35}$ European Organization for Nuclear Research (CERN), Geneva, Switzerland\\
% ...
$^{55}$ INFN, Sezione di Bologna, Bologna, Italy\\
% ...
$^{75}$ Johann-Wolfgang-Goethe Universit\"{a}t Frankfurt Institut f\"{u}r Informatik, Fachbereich Informatik und Mathematik, Frankfurt, Germany\\
% ...
$^{77}$ Korea Institute of Science and Technology Information, Daejeon, Republic of Korea\\
% ...
$^{82}$ Lund University Department of Physics, Division of Particle Physics, Lund, Sweden\\
% ...
$^{97}$ Nuclear Physics Institute of the Czech Academy of Sciences, \v{R}e\v{z} u Prahy, Czech Republic\\
% ...
$^{142}$ Variable Energy Cyclotron Centre, Homi Bhabha National Institute, Kolkata, India\\
% ...

\end{flushleft} 